\documentclass[journal=jctc,manuscript=article]{achemso}
\usepackage{chemformula} 
\ProvideTextCommand{\DJ}{OT1}{\leavevmode\raisebox{-.5ex}{\makebox[0pt][l]{\hskip-.07em\accent16\hss}}D}
\usepackage{adjustbox}
\usepackage{caption}
\usepackage{subcaption}
\usepackage{color}
\usepackage{hyperref}
\usepackage{booktabs}
\usepackage{multirow}

\usepackage[flushleft]{threeparttable}
\usepackage{booktabs,caption}
\usepackage{amsmath}
\usepackage{makecell}
\usepackage[table]{xcolor}
\usepackage{placeins}

\author{Xiangru Wang}
\affiliation[CUPB]{State Key Laboratory of Heavy Oil Processing, China University of Petroleum (Beijing), Beijing 102249, China}
\altaffiliation{Contributed equally to this work}
\author{Zekun Jiang}
\affiliation[CUPB]{State Key Laboratory of Heavy Oil Processing, China University of Petroleum (Beijing), Beijing 102249, China}
\altaffiliation{Contributed equally to this work}
\author{Heng Yang}
\affiliation[CUPB]{State Key Laboratory of Heavy Oil Processing, China University of Petroleum (Beijing), Beijing 102249, China}
\author{Cheng Tan}
\affiliation[CUPB]{Shanghai AI Laboratory,L1 Building, International Media Port, No. 129 Longwen Road, Xuhui District, Shanghai}
\author{Xingying Lan}
\affiliation[CUPB]{State Key Laboratory of Heavy Oil Processing, China University of Petroleum (Beijing), Beijing 102249, China}
\author{Chunming Xu}
\affiliation[CUPB]{State Key Laboratory of Heavy Oil Processing, China University of Petroleum (Beijing), Beijing 102249, China}

\author{Tianhang Zhou}
\affiliation[CUPB]{State Key Laboratory of Heavy Oil Processing, China University of Petroleum (Beijing), Beijing 102249, China}
\affiliation[CUPB]{College of Energy Innovation, China University of Petroleum (Beijing), Beijing 102249, China}
\email{zhouth@cup.edu.cn}

\title[An \textsf{achemso} demo]
  {Socrates-Mol: Self-Oriented Cognitive Reasoning through Autonomous Trial-and-Error with Empirical-Bayesian  Screening for Molecules}

\abbreviations{IR,NMR,UV}
\keywords{American Chemical Society, \LaTeX}

\DeclareUnicodeCharacter{2212}{-}

\usepackage{booktabs}
\usepackage{siunitx} 

\begin{document}
\pagebreak
 
\begin{abstract} 
Molecular property prediction is fundamental to chemical engineering applications such as solvent screening. We present Socrates-Mol, a framework that transforms language models into empirical Bayesian reasoners through context engineering, addressing cold start problems without model fine-tuning. The system implements a reflective-prediction cycle where initial outputs serve as priors, retrieved molecular cases provide evidence, and refined predictions form posteriors—extracting reusable chemical rules from sparse data. We introduce ranking tasks aligned with industrial screening priorities and employ cross-model self-consistency across five language models to reduce variance. Experiments on amine solvent LogP prediction reveal task-dependent patterns: regression achieves 72\% MAE reduction and 112\% R-squared improvement through self-consistency, while ranking tasks show limited gains due to systematic multi-model biases. The framework reduces deployment costs by over 70\% compared to full fine-tuning, providing a scalable solution for molecular property prediction while elucidating the task-adaptive nature of self-consistency mechanisms.
\end{abstract}

\pagebreak



\maketitle

\section{Introduction}\label{sec1}
Molecular property prediction is a critical step in chemical research and development, enabling the early evaluation of the physicochemical properties of compounds through computational methods. This predictive capability can reduce experimental trial-and-error costs and shorten development cycles in scenarios such as solvent screening and material design. In particular, accurate molecular property prediction accelerates the commercialization of new products, driving process optimization and industrial innovation in key areas, including high-performance solvent development and polymer modification. However, as chemical research advances toward complex molecular systems and novel compound design, the technical bottlenecks of existing prediction methods have become increasingly prominent, demanding urgent solutions\cite{zheng2025llm4sd, gervasio2024accelerating, mannhold2009calculation}.

From a methodological perspective, molecular property prediction in chemistry can be broadly categorized into three types. The first type comprises traditional machine learning models, such as support vector machines and random forests, which rely on manually designed molecular descriptors\cite{svetnik2003random, burbidge2001drug}. While these methods achieve stable performance in relatively simple molecular systems, their generalization ability degrades with increasing molecular structural complexity. Moreover, in data-scarce scenarios, they struggle to overcome the cold start problem and fail to function effectively. The second type involves physics-based molecular simulation methods, including molecular dynamics (MD) and density functional theory (DFT) calculations\cite{cohen2012insights, hollingsworth2018molecular}. These approaches can predict molecular behavior with high precision, making them valuable for analyzing complex molecular systems and capturing fine intermolecular interactions. However, their high computational costs limit the application in large-scale datasets and high-throughput screening tasks. The third type encompasses techniques based on large language models (LLMs). In recent years, specialized models like ChemBERTa, pretrained on chemical data, have enabled end-to-end prediction directly from SMILES strings, while general-purpose LLMs such as the GPT series have also shown potential in processing SMILES representations and reasoning about molecular properties\cite{ross2022large, weininger1988smiles, jablonka2024leveraging, ramos2024review}.

Despite their promising prospects across multiple domains, LLMs face significant challenges in chemical applications. On one hand, these models typically lack an inherent understanding of the underlying chemical principles, leading to mismatches between predictions and practical chemical \cite{nguyen2024can}. On the other hand, existing LLM-based research primarily focuses on regression tasks, whereas industrial practice places greater emphasis on the relative ranking of molecular properties (e.g., screening the top 30\% most hydrophobic solvents)\cite{li2025mppreasoner, zhang2025effective}. Thus, efforts solely aimed at improving regression accuracy may not directly address the needs of efficient molecular screening.

Academia has explored multiple directions to tackle these issues, but three key research gaps remain. First, regarding the cold start problem, most existing methods employ static data reuse strategies. For example, MolRAG constructs molecular retrieval libraries to assist LLM reasoning; however, such methods essentially treat data as passive indices and fail to extract reusable chemical knowledge, limiting their generalization across multi-property prediction tasks\cite{xian-etal-2025-molrag}. Similarly, ChemPrompt guides LLMs in learning chemical knowledge through domain-specific prompt engineering but relies heavily on large-scale labeled data, conflicting with the demands of low-data scenarios. Second, in terms of task adaptability, research on ranking tasks in chemistry is severely inadequate, lacking specialized reasoning mechanisms tailored to such tasks. Third, concerning result reliability, existing studies have proposed self-consistency mechanisms to mitigate LLM reasoning biases through multi-model collaboration; however, these mechanisms are not customized to the distinct characteristics of regression (emphasizing numerical precision) and ranking (emphasizing relative order), nor do they account for potential divergence in optimization effects caused by inherent task differences, thereby limiting their effectiveness in chemical applications\cite{gilmer2017neural, hu2024strategies}
.

In summary, this study proposes the Socrates-Mol framework. Its core idea is to extend LLM prediction behavior from one-time output to a multi-round dynamic reasoning process, enabling the model to self-calibrate through cycles of generation and reflection. Specifically: (1) A reflection-prediction loop is introduced to construct a self-reflective corpus, extracting reusable chemical rules from limited data to alleviate the cold start problem; (2) Ranking tasks are integrated into a unified framework to enhance adaptability in chemical screening scenarios; (3) Differentiated self-consistency mechanisms are designed for regression and ranking tasks. In structural design, Socrates-Mol functions not only as a multi-stage prediction system but also as an empirical Bayesian cognitive reasoner: the model's initial prediction corresponds to a prior belief, reflection and case retrieval provide empirical evidence for bias correction, and the self-consistency mechanism enables aggregation and rebalancing of multi-model posterior beliefs. Through these innovations, this research aims to establish a new paradigm for molecular property prediction, balancing accuracy, practicality, and generalization.

\begin{figure}[H]
\centering

\includegraphics[width=0.95\textwidth, height=0.5\textheight, keepaspectratio=false]{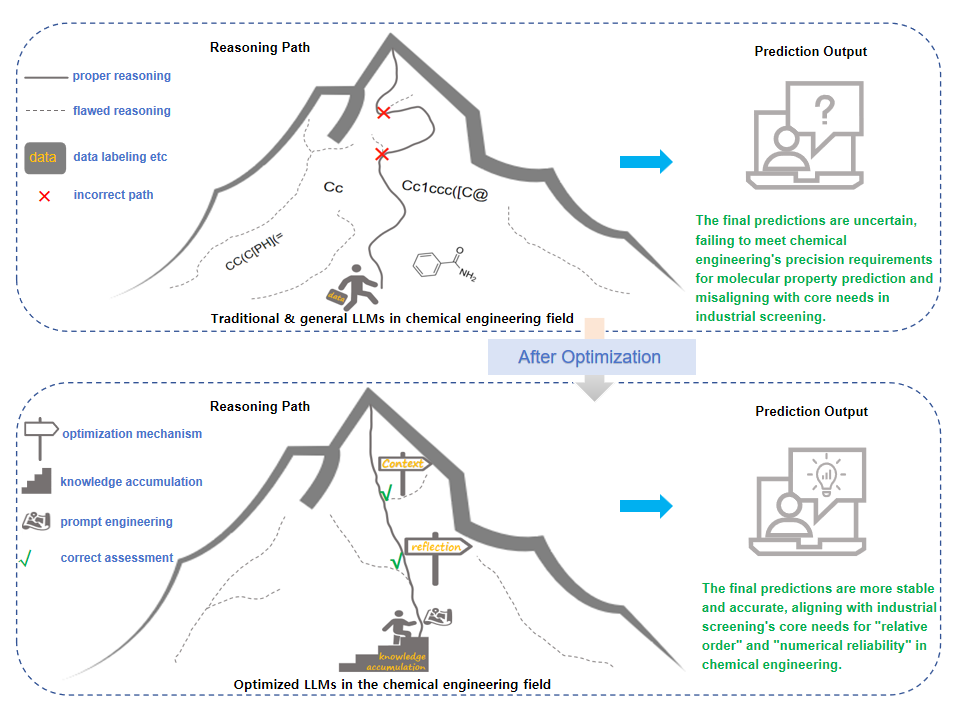}  
\includegraphics[width=1.0\textwidth]{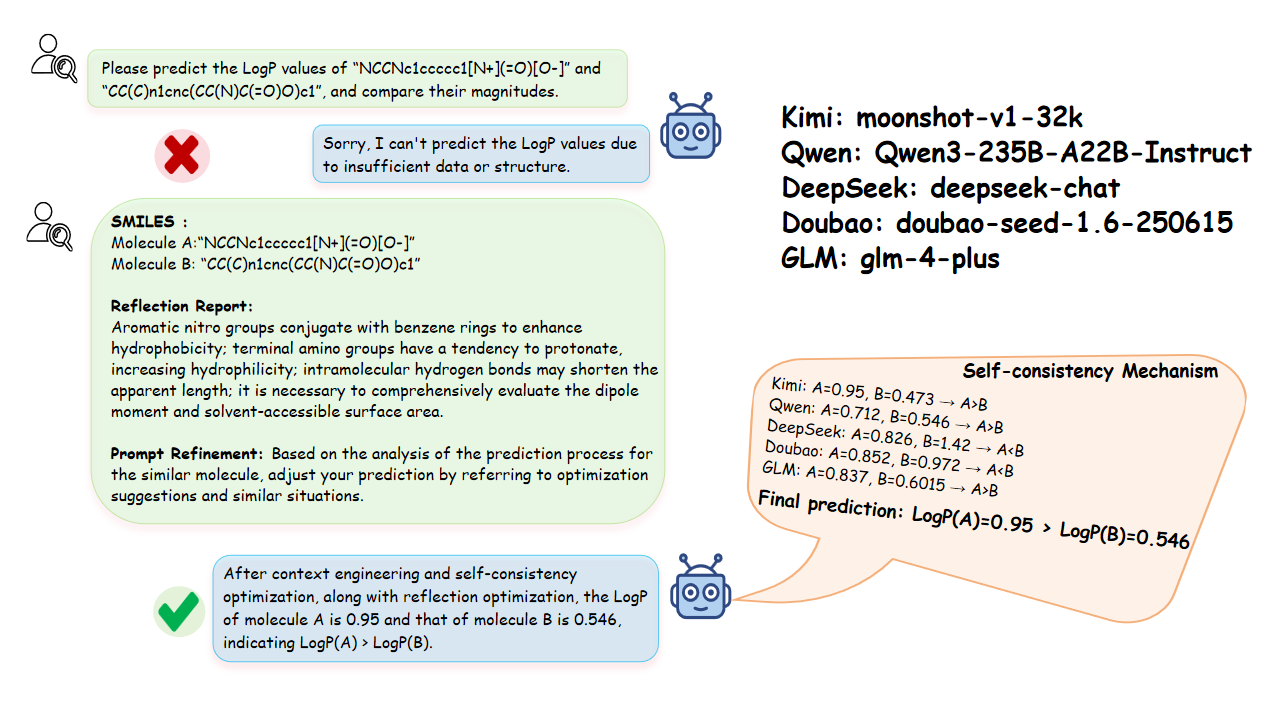}
\caption{A schematic of iterative reflection–prediction with self-consistency, illustrating how similar molecular cases and reflection-driven prompt optimization refine initially infeasible LogP predictions, which are subsequently stabilized by a self-consistency mechanism that consolidates multi-model outputs through task-specific consistency rules to yield robust final results.}
\label{fig:figure1}
\end{figure}

\section{Model and Method}

\subsection{Basic Setup}

This study selected a dataset containing 87 amine solvents, covering various structural types such as branched amines and heterocyclic amines to ensure the diversity of molecular structures in the dataset. Five representative general-purpose large language models were employed in the experiments: Kimi (moonshot-v1-32k, Moonshot AI), Qwen (Qwen3-235B-A22B-Instruct, Alibaba), DeepSeek (deepseek-chat, DeepSeek AI), Doubao (doubao-seed-1.6-250615, ByteDance), and GLM (glm-4-plus, Zhipu AI). These models represent the current state-of-the-art in Chinese language model development, each bringing distinct architectural characteristics and training paradigms to the ensemble. To reduce random errors and ensure statistical robustness, each model was run multiple times across all experimental conditions. The framework integrates four key components: context engineering, two-stage reflection mechanism, dynamic accumulation of real cases, and self-consistency mechanism. Among them, the self-consistency mechanism constitutes the core innovation of this study, designing differentiated execution strategies according to the distinct requirements of regression and ranking tasks to optimize posterior aggregation under the empirical Bayesian framework\cite{efron2012large, griffiths2020constrained}
.

\begin{figure}[H]
\centering
\includegraphics[width=0.95\textwidth, height=0.48\textheight, keepaspectratio=false]{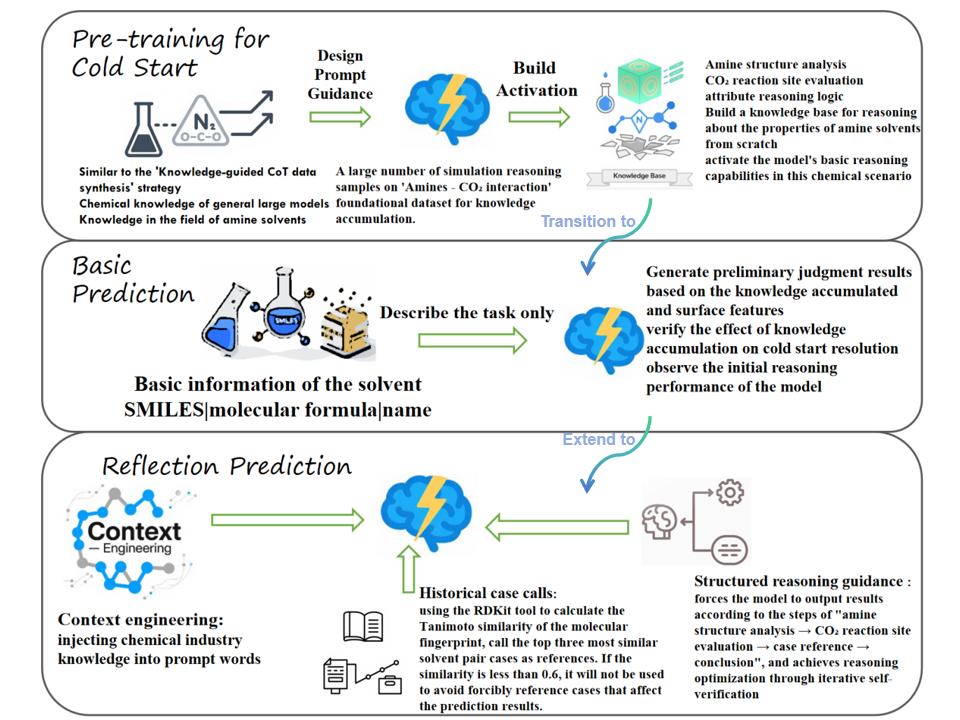}  
\hspace*{0.005\textwidth} 
\includegraphics[width=1.0\textwidth]{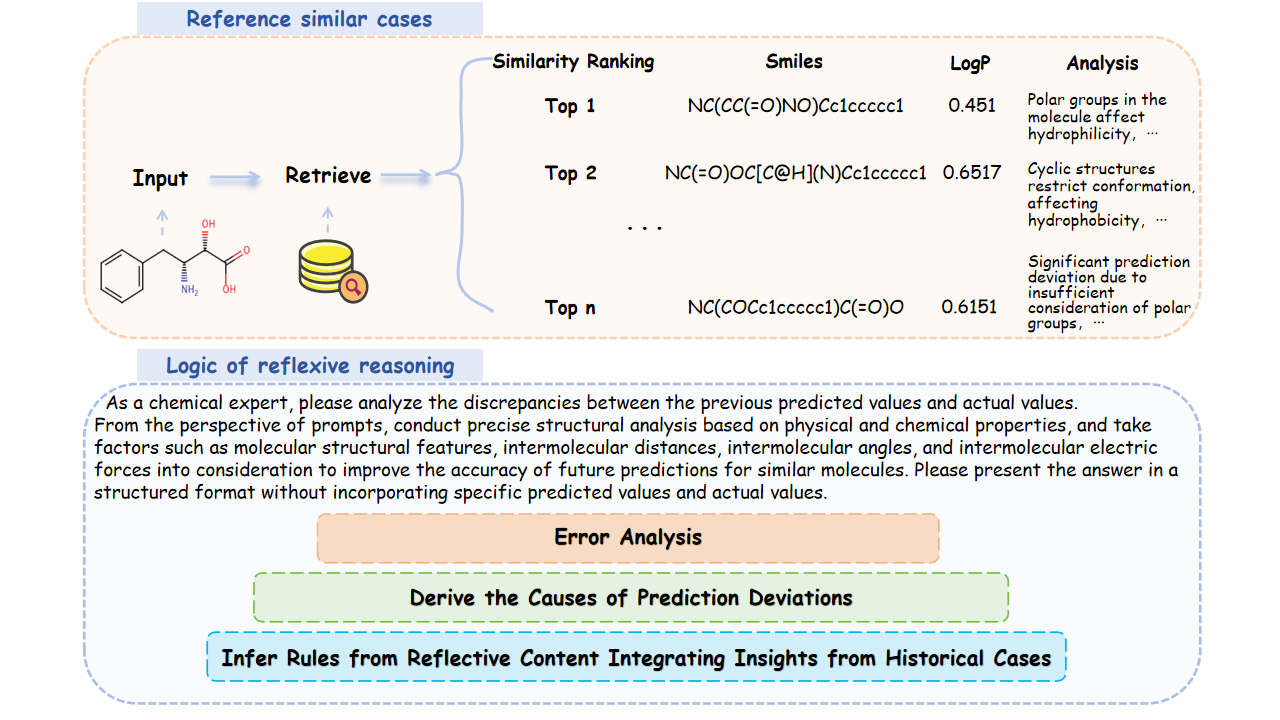}
\caption{A schematic illustrating the retrieval of structurally similar molecular cases and the logic of reflexive reasoning. Similarity-ranked reference molecules provide experimental LogP values and structural analyses, which guide error identification and interpretation. The reflective reasoning process analyzes prediction deviations, derives underlying causes, and formulates corrective rules informed by historical cases to improve future predictions.}
\label{fig:figure2}
\end{figure}


\subsection{Regression Task}
In this study, the main objective of the regression task was to leverage large language models to predict the LogP values of amine molecules while outputting the entire reasoning process. Within an empirical Bayesian framework, this process is conceptualized as a three-stage prior–evidence–posterior iterative cycle. First, the 87 amine solvents were sequentially introduced into five general purpose large language models, each of which was executed multiple times to reduce random variability. Using only SMILES representations and internal knowledge, the models generated initial LogP estimates, which were treated as the prior distribution $p(P \mid M)$.

Subsequently, the system compared these priors against experimental ground truths and automatically produced a difference analysis report, recording potential error sources such as scaffold differences, the number of polar functional groups, and stereochemical configurations as likelihood information. Similarly, molecules with a Tanimoto similarity greater than 0.1 were retrieved from the historical database, and their experimental values and reflective reports were incorporated as observed evidence $E$\cite{bajusz2015tanimoto}
.

Through reflective prompts, these structured pieces of evidence were integrated into the model context, allowing a Bayesian update to generate optimized posterior predictions $p(P \mid M, E)$. This mechanism preserves both the numerical results and the reasoning pathways associated with errors, gradually constructing a dynamically updated reflective case library to provide continuously refined priors for subsequent predictions.

\begin{figure}[H]
\centering
\includegraphics[width=1.0\textwidth]{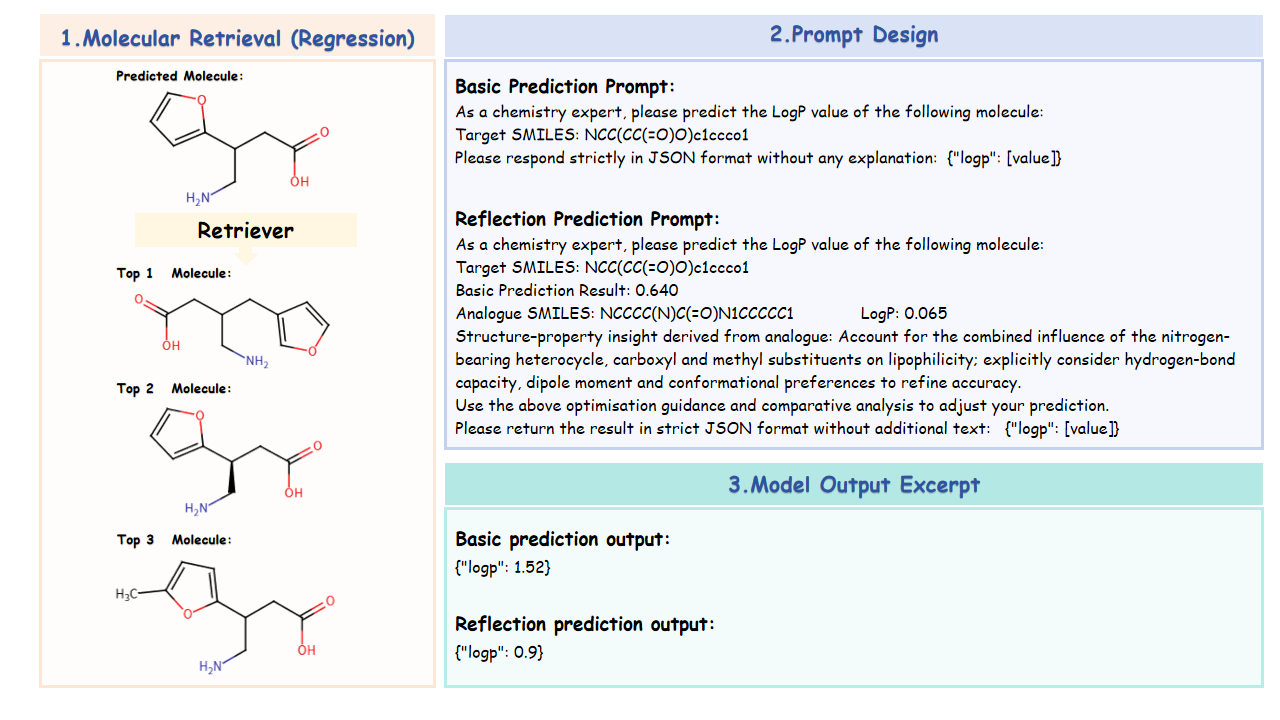}
\caption{A schematic diagram illustrating the LogP prediction workflow for regression tasks. It demonstrates how similar molecules are retrieved to provide a basis for prediction, how LogP prediction is guided by basic and reflection-enhanced prompts, and how prediction accuracy is improved through a reflection-based strategy.}
\label{fig:figure3}
\end{figure}

The experimental results demonstrate a significant reduction in prediction errors after posterior updating. However, in certain cases, the predicted results still exhibit instability, indicating that relying solely on numerical regression may not fully meet the stringent accuracy requirements of chemical engineering applications.

To evaluate the predictive accuracy of the regression task, the mean absolute error (MAE) was used as the loss function, defined as:
\begin{equation}
\text{MAE} = \frac{1}{n} \sum_{i=1}^{n} \left| y_{\text{true},i} - y_{\text{pred},i} \right|
\end{equation}

\noindent
where $y_{\text{true},i}$ denotes the ground-truth LogP value of the $i$-th sample, $y_{\text{pred},i}$ represents the corresponding predicted value, and $n$ is the total number of samples.

\begin{equation}
R^2 = 1 - \frac{\sum_{i=1}^{n} \left( y_{\text{true},i} - y_{\text{pred},i} \right)^2}{\sum_{i=1}^{n} \left( y_{\text{true},i} - \bar{y}_{\text{true}} \right)^2}
\end{equation}

\noindent
where $\bar{y}_{\text{true}}$ is the mean of the true values. An $R^2$ value closer to 1 indicates a better fit of the regression model.

\subsection{Ranking Task }
Given the limited predictive accuracy observed in the regression task and the emphasis on relative property comparison in chemical solvent screening, this study further introduces a ranking task. The primary objective of this task is to determine the relative magnitude of LogP values between two amine molecules while providing an interpretable reasoning process\cite{wang2024molecular}
.

The experiments used the same set of 87 amine solvents, with 80 pairs randomly sampled in each round. The five large language models were executed in parallel over multiple runs to reduce random variability. The prediction target was shifted from a single-point LogP estimate to determining the relative order while providing interpretable reasoning. Initially, the models generated prior beliefs $p(A > B \mid M)$ based solely on surface-level information, such as SMILES representations, molecular weight, and topological polar surface area, producing an initial judgment of either $A > B$ or $B > A$.

Subsequently, the system retrieved similar pairs from the historical database with Tanimoto similarity greater than 0.6. If a match was found, the experimental LogP difference along with reflective records was incorporated as observed evidence $E$. Using reflective prompts, the “difference direction + error explanation” was encoded as likelihood weights, enabling a Bayesian update to produce posterior ratios $p(A > B \mid M, E)$. If no sufficiently similar pair was found, chemical heuristics—such as “polar substituents decrease LogP” or “branching increases hydrophobicity”—were used to construct likelihoods, which were similarly incorporated into reflective prompts to correct the prior.

This dual strategy of case-driven reasoning and chemical-rule augmentation effectively mitigates prior biases across varying data densities. As a result, posterior ranking accuracy was significantly improved compared to the baseline stage, demonstrating robust performance for molecules with complex scaffolds or small $\Delta \text{LogP}$ differences. This approach better aligns with industrial requirements for solvent screening and materials optimization, where relative order is the key criterion.

\begin{figure}[H]
\centering
\includegraphics[width=1.0\textwidth]{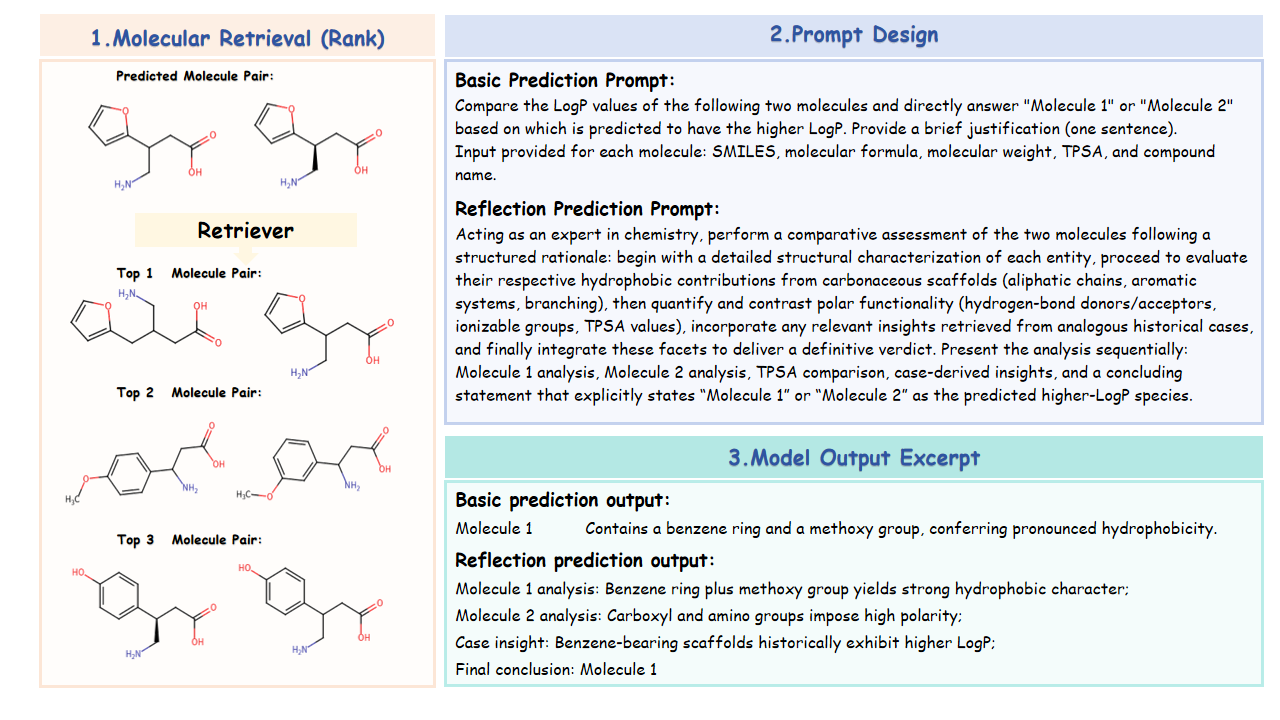}
\caption{A schematic diagram illustrating the LogP prediction process for ranking tasks. It demonstrates how similar molecule pairs are retrieved to provide a basis for prediction, how the LogP comparison prediction of molecule pairs is guided by basic and reflection-enhanced prompts, and how prediction accuracy is improved through a reflection-based strategy.}
\label{fig:figure4}
\end{figure}

Experimental results demonstrate that the ranking task substantially improved predictive accuracy and stability, particularly in the discrimination of complex molecular structures and subtle LogP differences. Compared with direct numerical regression, the ranking approach more effectively captures relative differences between molecules, thereby demonstrating better suitability and practical relevance for real-world applications such as solvent screening and materials optimization in chemical engineering.

To evaluate the performance of the ranking task, accuracy and F1-score were adopted as metrics. Accuracy is defined as:

\begin{equation}
\text{Accuracy} = \frac{\text{Number of correct rankings}}{\text{Total number of molecular pairs}}
\end{equation}

Here, ``Number of correct rankings'' denotes the molecular pairs for which the predicted LogP ordering matches the ground truth, and ``Total number of molecular pairs'' refers to the overall number of pairs involved in the experiments.

The F1-score, which balances precision and recall, is defined as:
\begin{equation}
\text{F1-score} = 2 \times \frac{\text{Precision} \times \text{Recall}}{\text{Precision} + \text{Recall}}
\end{equation}

where precision is the proportion of correctly predicted positive pairs among all predicted positive pairs (\( \text{Precision} = \frac{\text{True Positives}}{\text{True Positives} + \text{False Positives}} \)), and recall is the proportion of correctly predicted positive pairs among all actual positive pairs (\( \text{Recall} = \frac{\text{True Positives}}{\text{True Positives} + \text{False Negatives}} \)). For the ranking task, a ``positive pair'' is defined as a molecular pair where the ground truth satisfies \( \Delta \text{LogP}_{AB} > 0 \) (i.e., Molecule A has a larger LogP than Molecule B).







\subsection{Self-Consistency Mechanism}
The self-consistency mechanism is one of the core innovations of this study, aiming to improve the accuracy and robustness of both regression and ranking tasks through multi-model integrated prediction and voting strategies\cite{wang2023self, wei2022chain}
. Unlike traditional single-model prediction, this mechanism can effectively reduce the uncertainty of individual models, thereby achieving more stable results—particularly when handling complex molecular structures and boundary cases. Within the empirical Bayesian framework, the self-consistency mechanism is defined as a multi-posterior aggregation step, with its design fully preserving the task-specific differences between regression and ranking.

\subsubsection{Design of Self-Consistency Strategy for Regression Tasks}
For regression tasks, the self-consistency mechanism adopts a multi-model mean aggregation strategy, with the specific process as follows: The five LLMs first independently complete the aforementioned prior-evidence-posterior reasoning cycle, generating five differentiated posterior point estimates. Subsequently, the arithmetic mean of these five posterior values is calculated and adopted as the final output.

The core logic of this strategy lies in leveraging the mean convergence property of the ensemble posterior—since numerical biases in single-model regression tasks are more likely to manifest as random fluctuations, multi-model mean aggregation can effectively offset individual biases, driving results to converge toward the true value. Experimental validation demonstrates that this strategy achieves accuracy improvements through simple posterior mean calculation without introducing complex weighting rules or additional training parameters, which is highly aligned with the core requirement of regression tasks—pursuing stability and precision in numerical prediction—while balancing computational efficiency and result reliability.The formal definition is given as:

\begin{equation}
\hat{y}_{\text{final}} = \frac{1}{m} \sum_{i=1}^{m} \hat{y}_i
\label{eq:optimal_prediction}
\end{equation}

where  \( \hat{y}_i \) represents the posterior point estimate of the \( i \)-th model after reflection and case retrieval. This formula formalizes the multi-model mean aggregation strategy, aiming to offset random numerical biases through mean convergence.

\begin{figure}[H]
\centering
\includegraphics[width=1.0\textwidth]{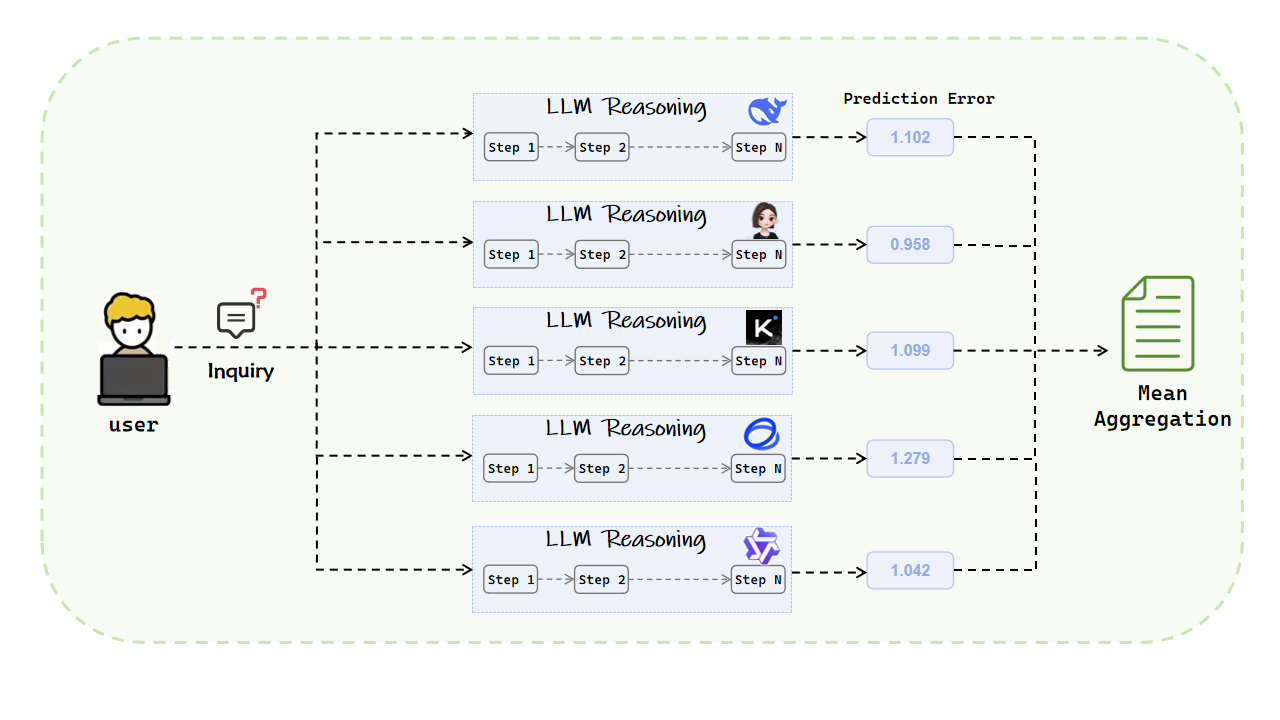}
\caption{A schematic of the self-consistency mechanism for regression tasks, illustrating how multiple LLMs independently perform reflective reasoning to generate LogP predictions, followed by a mean aggregation process that computes the arithmetic average of these predictions to ensure consistency and robustness across models.}
\label{fig:figure5}
\end{figure}

\subsubsection{Design and Optimization of Self-Consistency Strategy for Ranking Tasks}
Ranking tasks require a focus on the reliability of relative order of molecular properties, but a single aggregation strategy may fail to address potential cognitive biases across multiple models\cite{ju2024multi}
. Therefore, we designed and validated multiple sets of self-consistency schemes based on different integration logics. All schemes were tested on the same batch of amine solvent molecule pairs, with the basic prior-evidence-posterior reasoning process kept consistent; only the aggregation method of multi-model posteriors was altered to ensure that performance differences could be attributed to the integration logic itself. Specifically, we first attempted to leverage the chemical reasoning advantages demonstrated by the Doubao model in previous single-task experiments, designating it as the final reviewer to integrate reflection results from the five models and output the final ranking through multi-source information synthesis. Considering the risk of information overload in single-model integration, we then formed an expert panel consisting of three models (Doubao, Qwen, and GLM) with richer pre-training data in the chemical domain, which made voting decisions after independently evaluating all models’ reflection results. Meanwhile, to compare the performance of baseline and optimized schemes, we also tested the traditional majority voting strategy, directly determining results based on the ranking direction of the five models. Finally, to address the issue that baseline strategies may overlook differences in model performance, we designed a dynamic weight adjustment scheme, which real-time assigns voting weights based on each model’s historical ranking accuracy (e.g., a weight of $1.2$ for accuracy $\geq 85\%$ and $0.6$ for $< 70\%$). Through weighted voting to output results, this scheme aims to strengthen the decision-making contribution of high-performance models.The formal definitions corresponding to each strategy are expressed as:

\begin{equation}
\hat{y}_{\text{final}} = f_{\text{Doubao}} \left( \{\hat{y}_1, \dots, \hat{y}_m, \mathcal{R}_1, \dots, \mathcal{R}_m\} \right)
\label{eq:single_model_integration}
\end{equation}

\begin{equation}
\hat{y}_{\text{final}} = \arg \max \left( \sum_{k \in \text{Panel}} \hat{y}_k \right), \quad \text{Panel} = \{\text{Doubao}, \text{Qwen}, \text{GLM}\}
\label{eq:expert_panel_voting}
\end{equation}

\begin{equation}
\hat{y}_{\text{final}} = \arg \max \left( \sum_{i=1}^{m} \hat{y}_i \right), \quad \hat{y}_i \in \{+1, -1\}, \, m=5
\label{eq:majority_voting}
\end{equation}

\begin{equation}
\hat{y}_{\text{final}} = \arg \max \left( \sum_{i=1}^{m} w_i \cdot \hat{y}_i \right), \quad w_i = f(\text{Acc}_i)
\label{eq:dynamic_weight_voting}
\end{equation}

where $\hat{y}_i \in \{+1, -1\}$ denotes the ranking decision of the $i$-th model ($+1$ indicates that molecule A is superior to molecule B in terms of molecular properties, and $-1$ otherwise); $w_i$ represents the dynamic weight of the $i$-th model, determined by its historical ranking accuracy through the mapping function $f(\text{Acc}_i)$ (e.g., $w_i=1.2$ if $\text{Acc}_i \geq 85\%$, and $w_i=0.6$ if $\text{Acc}_i < 70\%$). This formula formalizes the weighted voting logic, emphasizing the contribution of high-performance models in the final decision.

\begin{figure}[H]
\centering
\includegraphics[width=1.0\textwidth]{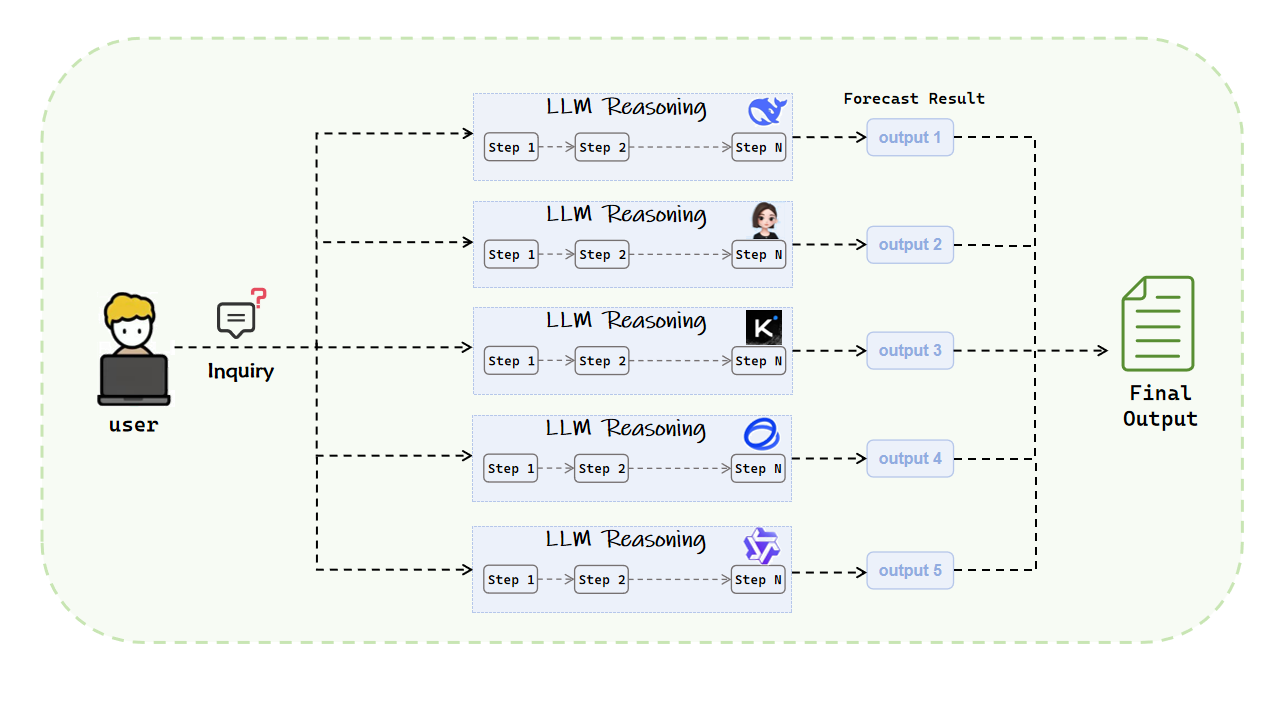}
\caption{A schematic of the self-consistency mechanism for ranking tasks, showing how multiple LLMs independently conduct pairwise comparative reasoning, and the final decision is determined through a majority voting strategy that selects the most consistent comparative outcome, ensuring stability and reliability across models.}
\label{fig:figure6}
\end{figure}

Experimental results indicate that the self-consistency mechanism significantly improves both the accuracy and stability of predictions in regression and ranking tasks. More importantly, in chemical engineering applications, this mechanism provides reliable and precise support for tasks such as rapid solvent screening and materials optimization, thereby underscoring its practical value.

\section{Results and Discussions}
\subsection{Regression Task Results}
Experimental results indicate that baseline regression predictions have significant limitations in terms of accuracy. However, with the introduction of the framework proposed in this study, model performance has improved significantly: according to the mean absolute error (MAE) evaluation, after reflection-based prediction, the MAE decreased from 0.86 to 0.62502; meanwhile, the coefficient of determination \( R^2 \) increased from -7.80348 to -4.06112. These results demonstrate that the framework effectively integrates historical case knowledge and discrepancy analysis, dynamically optimizes initial predictions, and thereby enhances the model's adaptability to complex molecular scenarios. Nevertheless, there remains a non-negligible gap between predicted and true values, highlighting the limitations of relying solely on numerical regression in scenarios with high precision requirements.

\begin{figure}[H]
\centering
\includegraphics[width=1.0\textwidth]{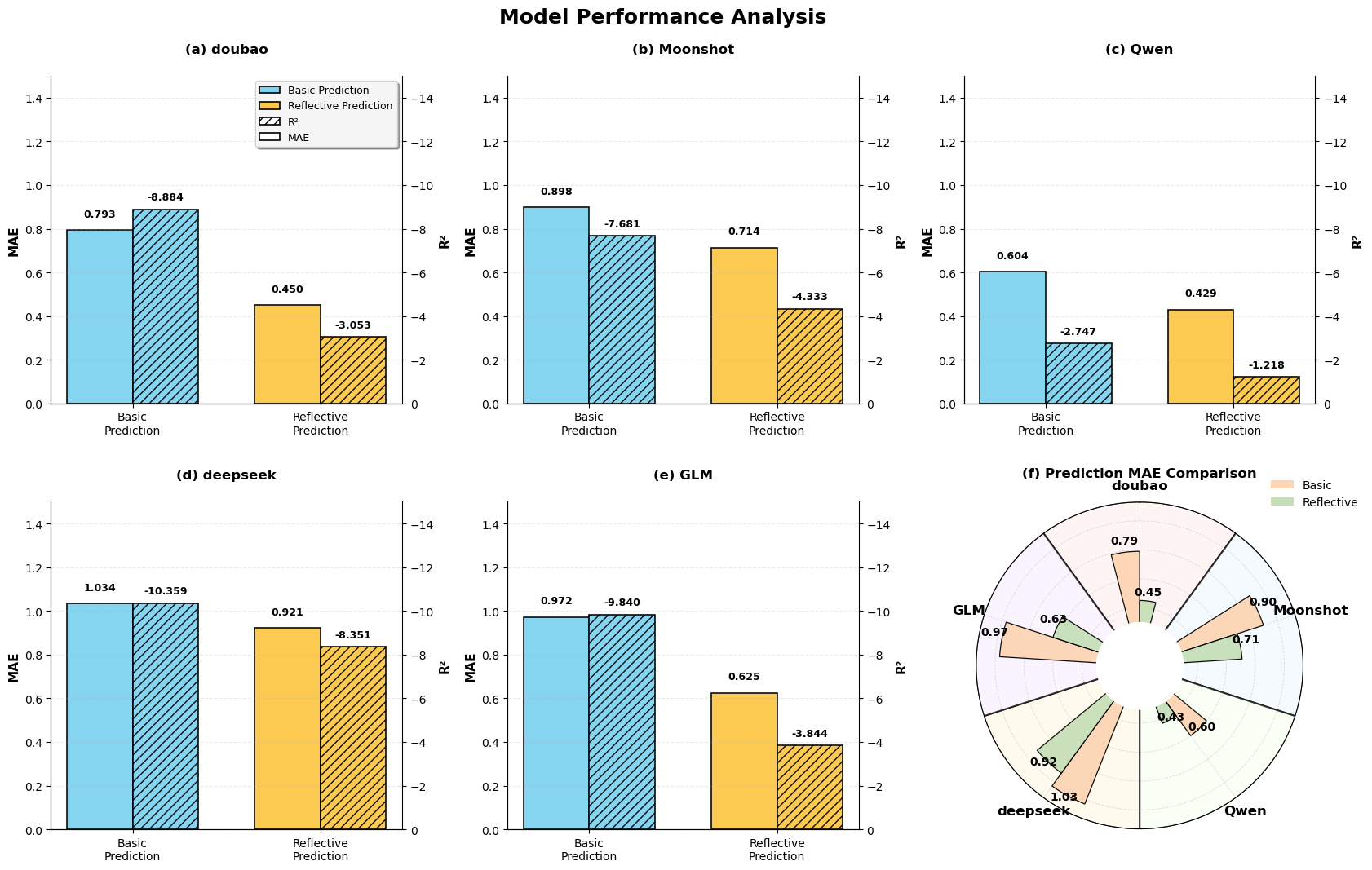}
\caption{This figure provides an intuitive visualization of the performance and horizontal comparison of the five models. Specifically, subfigures (a)-(e) present bar charts illustrating the MAE and $R^2$ of each model under the two strategies, while subfigure (f) displays a radial bar chart that comprehensively compares the two prediction strategies across all models from the perspective of MAE.}
\label{fig:figure3}
\end{figure}

\subsection{Ranking Task Results}
Experimental results further demonstrate that the overall accuracy of the ranking task is relatively high. Among the average results of the five models, the baseline prediction accuracy is 50\%, and after introducing the proposed framework, the accuracy increases to 71.8\%, with a relative improvement of 43.6\%. Additional analyses confirm that the model maintains stable performance across samples of varying difficulty, highlighting the robustness of the proposed framework in distinguishing differences in LogP values. Overall, the ranking task not only enhances the reliability of predictions but also exhibits strong applicability and practical value in chemical engineering scenarios, particularly in solvent screening and material optimization.

\begin{figure}[H]
\centering
\includegraphics[width=1.0\textwidth]{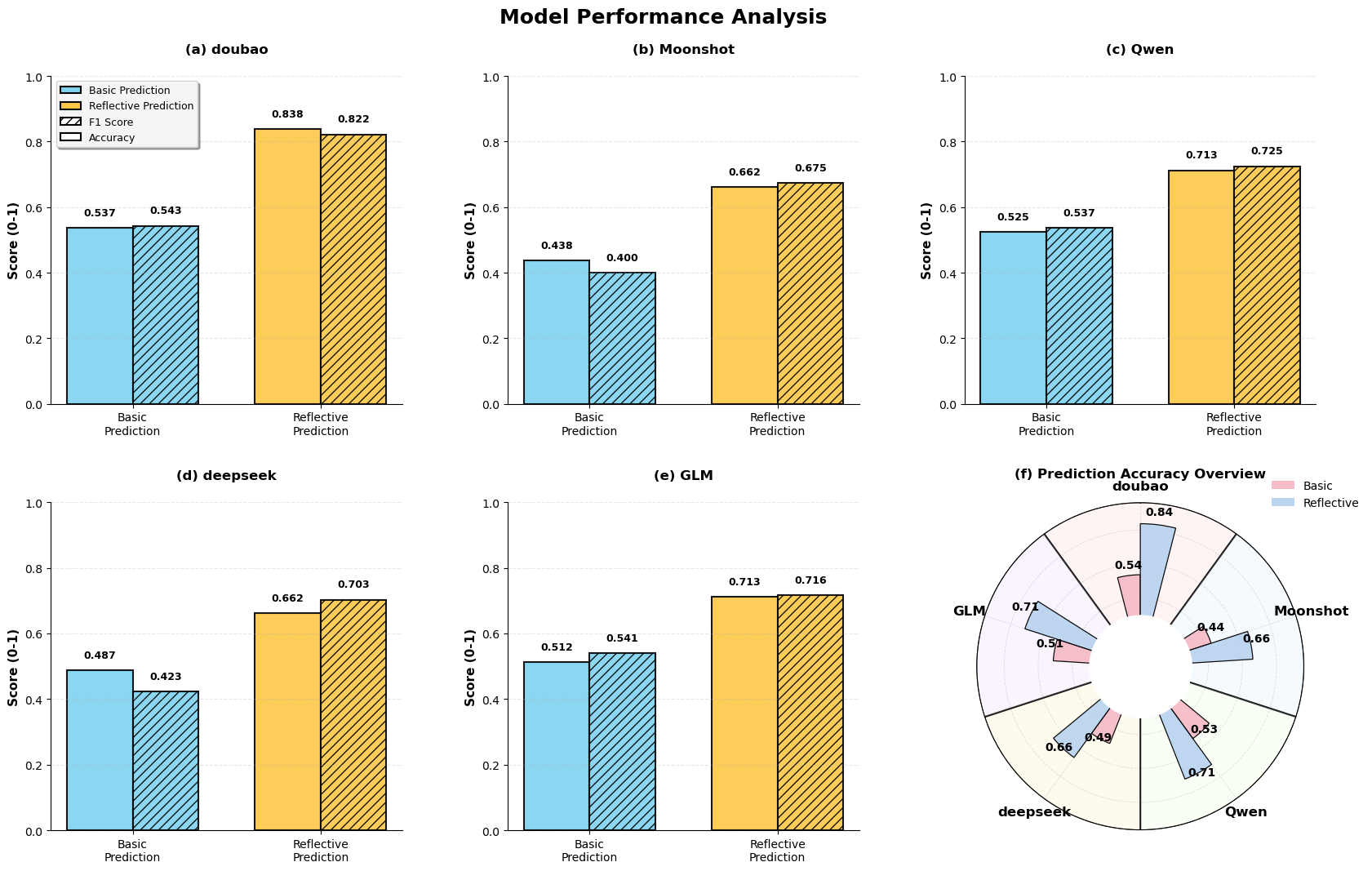}
\caption{This figure provides an intuitive visualization of the performance and horizontal comparison of the five models. Specifically, subfigures (a)-(e) present bar charts illustrating the Accuracy and F1 Score of each model under the two strategies, while subfigure (f) displays a radial bar chart that comprehensively compares the two prediction strategies across all models from the perspective of performance.}
\label{fig:figure3}
\end{figure}

\subsection{Results of the Self-Consistency Mechanism}
The experimental results reveal significant differences in the optimization effects of the self-consistency mechanism across different tasks: it demonstrates excellent capabilities in improving accuracy and stability for regression tasks. In contrast, for ranking tasks, despite our attempts to implement multiple sets of differentiated aggregation strategies, no significant performance breakthrough was achieved. Furthermore, a phenomenon emerged where the prediction accuracy of some models, after reflection, exceeded that of the self-consistency optimized results.

In the regression task,the average of the reflection-based prediction results from the five models shows a mean absolute error (MAE) of 0.62502 and a mean coefficient of determination ($R^2$) of -4.06112, indicating that significant numerical biases remain after single-model reflection. After introducing the multi-model mean aggregation self-consistency mechanism—where the arithmetic mean of the posterior point estimates from the five models is computed as the final output—the average MAE decreases to 0.2586, while the average $R^2$ improves to 0.1591. These results verify the high compatibility between regression tasks and the multi-model mean aggregation strategy: since numerical biases in single models are mostly manifested as random fluctuations, mean aggregation can effectively offset individual biases, driving results to converge toward the true values. This mechanism achieves simultaneous improvements in accuracy and stability through simple posterior mean calculation without requiring complex weighting rules or additional training parameters, which is highly aligned with the core requirement of regression tasks—pursuing reliability in numerical prediction—while maintaining computational efficiency.

\begin{table}[htbp]
  \centering
  \caption{Performance of Basic Prediction, Reflective Prediction and Self-Consistency Mechanism in Regression Task}
  \label{tab:regression_performance}
  \begin{tabular}{lcccc}
    \toprule
    \textbf{Model} & \makecell{\textbf{Basic}\\ \textbf{MAE}} & 
    \makecell{\textbf{Reflective}\\ \textbf{ MAE}} & 
    \makecell{\textbf{Basic}\\ \textbf{\( R^2 \)}} & 
    \makecell{\textbf{Reflective}\\ \textbf{ \( R^2 \)}} \\
    \midrule
    GLM      & 0.9618 & 0.6106 & -9.5366 & -3.5401 \\
    Deepseek & 0.7902 & 0.4432 & -8.7854 & -2.9541 \\
    Doubao   & 1.0403 & 0.9268 & -10.342 & -8.3343 \\
    Qwen     & 0.6098 & 0.4332 & -2.7465 & -1.2178 \\
    Kimi     & 0.8979 & 0.7113 & -7.609  & -4.2593 \\
    \midrule
    \makecell{\textbf{All Models}\\ \textbf{Average}} & 0.86 & 0.62502 & -7.80348 & -4.06112 \\
    \midrule
    \rowcolor{gray!18}
    \textbf{Self - Consistent Mechanism} & 0.6279 & 0.2586 & -4.1599 & 0.1591 \\
    \midrule
    \textbf{Relative Improvement} & \multicolumn{4}{c}{
      \makecell{
        Reflective Average MAE Reduction: 58.63\% \\ 
        Reflective Average \( R^2 \) Increase: 103.92\%
      }
    } \\
    \bottomrule
  \end{tabular}
\end{table}
\FloatBarrier

In the ranking task, none of the test results for the multiple self-consistency strategies met expectations. Initially, Doubao was tasked with independently integrating the reflection results from the other four models; however, its accuracy failed to improve consistently. Analysis revealed that Doubao failed to identify the common cognitive biases among other models and instead incorporated erroneous reasoning logic into decision-making, leading to additional integration errors. Subsequently, an expert panel consisting of Doubao, Qwen, and GLM was formed for voting. Although the prediction accuracy for some molecular pairs improved slightly, the three models shared similar blind spots in judging the hydrophobicity of hydroxyl and heterocyclic structures, resulting in synchronous misjudgments in cases such as hydroxyl-containing heterocyclic amines vs. hydroxyl-free linear amines. The traditional majority voting strategy yields even more limited effectiveness, and for molecular pairs with subtle differences, the misjudgment rate increases by 3.2\% instead. Finally, the tested dynamic weight adjustment scheme, despite its ability to update model weights based on accuracy, failed to break through the bottleneck due to the non-independent error distribution across multiple models. 

\begin{table}[htbp]
  \centering
  \caption{Performance of Basic Prediction, Reflective Prediction and Self-Consistency Mechanism in Partial Ranking Task}
  \label{tab:regression_performance}
  \begin{tabular}{lcccc}
    \toprule
    \textbf{Model} & \makecell{\textbf{Basic}\\ \textbf{Accuracy}} & 
    \makecell{\textbf{Reflective}\\ \textbf{Accuracy}} & 
    \makecell{\textbf{Basic}\\ $\mathbf{F1\text{-}score}$} & 
    \makecell{\textbf{Reflective}\\ $\mathbf{F1\text{-}score}$} \\
    \midrule
    GLM      & 38\% & 64\% & 0.4848 & 0.8095 \\
    Deepseek & 42\% & 64\% & 0.4231 & 0.6538 \\
    Doubao   & 48\% & 88\% & 0.7013 & 0.8235 \\
    Qwen     & 46\% & 68\% & 0.5512 & 0.6953 \\
    Kimi     & 44\% & 50\% & 0.5000  & 0.7308 \\
    \midrule
    \makecell{\textbf{All Models}\\ \textbf{Average}} & 43.6\% & 66.8\% & 0.53208 & 0.74258 \\
    \midrule
    \rowcolor{gray!18}
    \textbf{Self - Consistent Mechanism} & 42\% & 72\% & 0.6111 & 0.7500 \\
    \midrule
    \textbf{Relative Improvement} & \multicolumn{4}{c}{
      \makecell{
        Reflective Average Accuracy Increase: 7.78\% \\ 
        Reflective Average F1-score Increase: 1.00\%
      }
    } \\
    \bottomrule
  \end{tabular}
\end{table}
\FloatBarrier

The performance discrepancy between the two types of tasks stems from differences in task nature and bias characteristics: regression tasks are dominated by random errors, which can be quantified through absolute deviations to screen for optimal results; in contrast, ranking tasks are primarily affected by systematic biases. Coupled with insufficient valid evidence that renders biases difficult to correct through cases, existing aggregation strategies fail to fundamentally address this issue. This finding not only deepens our understanding of the limitations of self-consistency mechanisms but also provides a critical experimental foundation for designing bias-aware aggregation strategies for ranking tasks and improving evidence retrieval efficiency

\section{Conclusion}

The Socrates-Mol framework reconceptualizes molecular property prediction as empirical Bayesian inference\cite{gelman2013bayesian}, where context engineering reshapes language models into probabilistic reasoners operating through explicit prior-evidence-posterior cycles. Initial predictions establish priors $p(P|M)$ derived from model pretraining, reflective retrieval injects observational evidence $E$ from dynamic case libraries\cite{lewis2020retrieval}, and self-consistency aggregates multi-model posteriors\cite{wang2022self} $p(P|M,E)$ to complete uncertainty-reducing inference loops. Systematic evaluation across 87 amine solvents confirms that this Bayesian paradigm consistently outperforms baseline approaches under data-scarce and cold-start conditions, with pronounced advantages in resolving subtle structural differences and complex property landscapes.

Task-specific experiments reveal distinct optimization patterns across different prediction tasks. Regression tasks achieve substantial performance gains under self-consistency aggregation, with MAE reduced by approximately 72.42\% and $R^2$ improved by approximately 112.45\% relative to baseline performance. These improvements arise from the quantifiable nature of numerical biases, which facilitates effective posterior alignment across models. In contrast, ranking tasks exhibit superior chemical discrimination in their initial predictions and better alignment with industrial screening workflows that prioritize relative property ordering. However, they fail to achieve comparable optimization under consistency mechanisms, even after exploring various aggregation strategies including dynamic weighted voting and threshold correction. This asymmetry arises from systematic cross-model biases and insufficient retrieval of discriminative evidence in boundary cases, highlighting the task-dependent effectiveness of self-consistency as a posterior aggregation approach.

These findings validate three core principles underlying the framework. First, context engineering effectively guides models toward task-relevant reasoning trajectories\cite{wei2022chain}, demonstrating the utility of structured prompting in scientific domains\cite{white2023prompt}. Second, dynamic case libraries enable continuous knowledge accumulation through iterative reflection, forming self-improving prior distributions. Third, self-consistency mechanisms substantially reduce single-model variance in regression contexts while exposing fundamental limitations in ranking scenarios that demand methodological refinement.

Future development should proceed along three interconnected directions, with targeted enhancements addressing the identified bottlenecks in ranking tasks. First, integrating retrieval-augmented generation~\cite{lewis2020retrieval}, graph neural networks~\cite{gilmer2017neural}, or multimodal encoders to construct molecule-pair-specific indices and fine-grained structural feature extractors can improve evidence retrieval accuracy and bias correction while enriching the likelihood estimation $p(E|P)$ with stronger structural priors for complex chemical systems. Second, expanding multi-property case libraries with strategic emphasis on high-value ranking examples---such as molecular pairs with marginal LogP differences or complex skeletal variations---enables cross-task knowledge transfer and sustains a knowledge-reasoning feedback loop. Third, embedding the framework within high-throughput screening platforms through a two-stage architecture can directly support industrial decision-making while reducing research and development costs: regression-based coarse filtering leveraging self-consistency optimization, followed by ranking-based refinement incorporating bias correction.

At the application frontier, the framework extends naturally from solvent screening to polymer modification, reaction condition optimization, and broader material genome initiatives\cite{jain2013commentary}, providing interpretable decision support for fine chemical synthesis and material design. At the paradigm level, the reflection-driven Bayesian cycle transcends chemistry, offering transferable strategies for computation-intensive and knowledge-driven tasks across scientific domains. The prior-evidence-posterior update logic and task-adaptive optimization principles establish a general architecture for next-generation scientific computing tools that balance precision, robustness, and scalability\cite{brown2020language}.

Through continued methodological refinement and cross-domain exploration, we envision Socrates-Mol evolving into a comprehensive intelligent reasoning platform that accelerates dual transformations in industrial innovation and fundamental research\cite{schwaller2019molecular,zhou2023large}, demonstrating how empirical Bayesian principles can reshape how language models engage with scientific knowledge.

\bibliography{main}

\end{document}